\documentclass[aps,prl,floatfix,twocolumn,footinbib,groupedaddress]{revtex4-1}
\usepackage{epsfig}
\usepackage{graphicx}% Include figure files
\usepackage{dcolumn}   % needed for some tables
\usepackage{bm}% bold math
\usepackage{amssymb}
\usepackage{mathrsfs}
\usepackage{amsmath}
\usepackage{bbold}
\usepackage{epstopdf}

\usepackage[colorlinks=true,linkcolor=blue,anchorcolor=blue,citecolor=blue,urlcolor=blue]{hyperref}

\begin{document}

\title{Electroluminescence and thermal radiation from metallic carbon nanotubes with defects}

\preprint{APS/123-QED}
\author{Zu-Quan Zhang}
\email{phyzhaz@nus.edu.sg}
\author{Jian-Sheng Wang}

\affiliation{Department of Physics, National University of Singapore, Singapore 117551, Republic of Singapore}

\date{9 June 2021}

%\date{\today}
% It is always \today, today,
% but any date may be explicitly specified

\begin{abstract}
Bias-induced light emission and thermal radiation from conducting channels of carbon nanotubes (CNTs) with defects  are studied theoretically within the framework of nonequilibrium Green's function method based on a tight-binding model. Localized states induced by the single vacancy defect and single Stone-Wales defect in the low energy range enhance electroluminescence significantly while they reduce thermal radiation under zero bias. The influence of the diameters of the CNTs with defects on the radiation is discussed. Different from the 2D or bulk materials, the radiation intensities from quasi-one-dimensional metallic CNTs in thermal equilibrium are much smaller than that of the black-body radiation. We attribute this to the confinement of thermal excitation in the transverse direction of the CNT. Our study is important for optoelectronic applications of CNTs with defects.
\end{abstract}

\maketitle

%%%%%%%%%%%%%%%%%%%%%%%%%%%%%%%%%%%%%%%%%%%%%%
%\section{\label{sec:Intro} Introduction}
%%%%%%%%%%%%%%%%%%%%%%%%%%%%%%%%%%%%%%%%%%%%%%
{\emph{Introduction.}}
Single-walled carbon nanotubes ({SWCNTs}) are quasi-one-dimensional materials possessing extraordinary electrical, mechanical and optoelectronic properties~\cite{RevModPhys.79.677}. They can be metallic, small-gap semiconducting or semiconducting~\cite{ouyang2001energy, CaoJien2003}. The geometry of a SWCNT can be described by the tube's chiral vector which is defined by a pair of integers $(n, m)$~\cite{RevModPhys.79.677}. The band structure of a metallic SWCNT shows nearly linear dispersion relation in the low energy range around the Fermi level. Optical transitions are forbidden in this low energy range under an external electric field along the tube axis~\cite{Ando2005}. When the CNTs are stimulated by electrons or photons, significant light emissions can occur due to transitions between pairs of van Hove singularities that are mirror symmetric with respect to the Fermi level. 

Electroluminescence (EL) and photoluminescence (PL) from SWCNTs have been studied a lot in semiconducting CNTs experimentally~\cite{misewich2003electrically, chen2005bright, PhysRevLett.98.167406, avouris2008carbon}. In contrast, observations of EL from metallic SWCNTs are much less frequent~\cite{mann2007electrically, xie2009electroluminescence, essig2010phonon}.  Electroluminescence from suspended metallic SWCNTs are explained by Joule heating~\cite{mann2007electrically}, and the emission spectrum is different from black-body-like emissions discovered in nanotube bundles and multiwalled CNTs~\cite{Sveningsson2002, LiPeng2003, Jinquan2004}. The important role of phonons in light emission has been stressed in recent experiments, where a side peak close to the main transition peak due to phonon-assisted emission appears in the radiation spectrum~\cite{xie2009electroluminescence, essig2010phonon}. 

Defects in CNTs are widely studied and they are shown to have significant influences on various properties of CNTs, such as electric and magnetic properties~\cite{shtogun2009electronic, Partovi_Azar_2011, Jianhua2011},  transport properties~\cite{PhysRevB.54.2600, PhysRevLett.84.2917, Neophytou2007, TEICHERT201749}, field emission~\cite{WeiGu2006}, mechanical and optical properties~\cite{BUONGIORNONARDELLI20001703, PhysRevB.70.245416, SHARMA20123373, Harutyunyan2009, Jinglin2019}. Common atomic-scale defects in CNTs are vacancies, adatoms and Stone-Wales (SW) reconstruction~\cite{RevModPhys.79.677, fan2005identifying}. Recent experiments have shown that defects can be engineered to tune the  optic properties of the CNTs~\cite{brozena2019controlling}, such as enhancing the PL and tuning the single photon emission by $sp^{3}$ defects~\cite{piao2013brightening, he2017tunable, Ishii2018}. In contrast, the influence of defects on the EL from CNTs is challenging and less studied~\cite{coratger2001stm, UEMURA2006L15, Katano2018}. A recent experiment showed that a local defect could be induced by injecting tunneling electrons from a scanning tunneling microscope (STM) tip to a multiwalled CNT, and corresponding changes of EL due to the defect were observed~\cite{Katano2018}. Theoretically, bias-induced light emission from nanoscale system has received much attention in recent years~\cite{galperin2012, Jingtao2013, Kaasbjerg2015, Miwa2019, Parzefall_2019, Ridley2021}, especially in molecular junctions. However, few works have taken into account of the full geometry of the system at the atomic-scale level.  Quantitative calculations of EL from metallic SWCNTs and also taking into account of the influence of defects would be helpful to related experiments. 

In this work, we consider a two-terminal device to study the EL from the conducting channels of metallic SWCNTs under the influences of the single vacancy (SV) defect and the single SW defect. We consider electron transport in the ballistic regime and electron-phonon interaction is not included. This can be reasonable as the length of the conducting channel considered here is much smaller than the electron mean free path of the metallic SWCNT, which is about several micrometers~\cite{PhysRevLett.98.186808}. By turning off the applied bias in the device, we also study the thermal radiation from perfect and defected CNTs. 

\begin{figure} 
\centering
\includegraphics[width=8 cm]{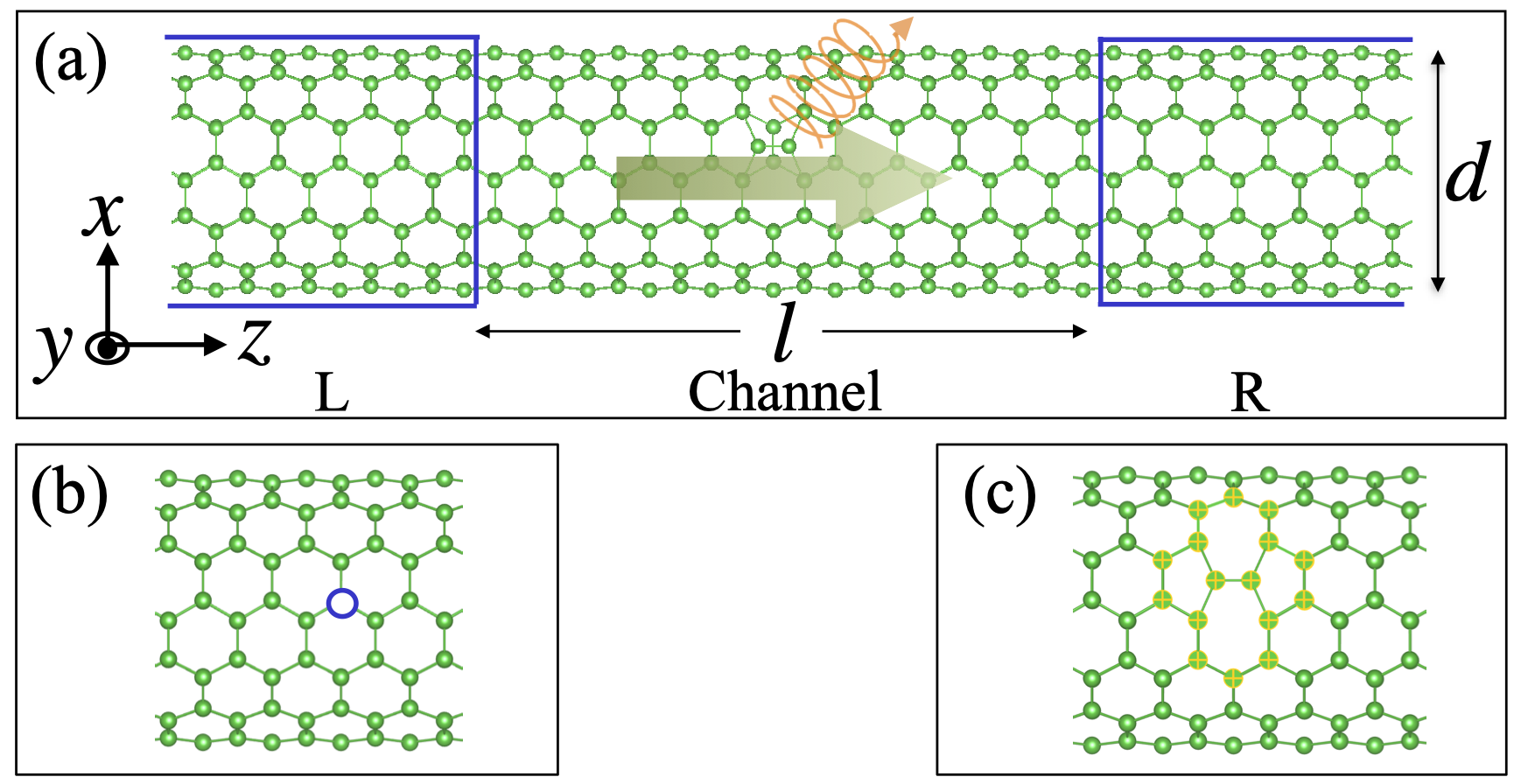}
\caption{Illustration of a two-terminal transport device of a SWCNT.  (a) The left (L) and right (R) leads are semi-infinite extensions of the pristine carbon nanotube with diameter $d$, and central region is the conducting channel with a finite length $l$. The interaction of electrons with the EM field is included only in the central region. The driving current in the central channel induces light emission.  (b) A single vacancy defect in the channel, with the missing atom denoted by a blue ring. (c) A Stone-Wales defect in the channel. The structures of nanotubes are drawn using VESTA 3~\cite{Momma2011}.
}
\label{fig:MD}
\end{figure}

%%%%%%%%%%%%%%%%%%%%%%%%%%%%%%%%%%%%%%%%%%%%%%
%\section{\label{sec:ModelAndMethod} Model and Theory}
%%%%%%%%%%%%%%%%%%%%%%%%%%%%%%%%%%%%%%%%%%%%%%
{\emph{ Theory.}}
We describe the Hamiltonian of the electrons in the CNT using the nearest-neighbor (NN) tight-binding (TB) model
\begin{equation} \label{eq:H0}
H_{0} = - \sum_{\langle ij \rangle}  t_{ij} c_{i}^{\dag} c_{j},
\end{equation}
where $t_{ij}$ is the hopping parameter, $c_{i}^{\dag}$ ($c_{j}$) is the electron creation (annihilation) operator on site $i$ (site $j$), the angular bracket $\langle ij \rangle$ denotes NN sites. We introduce the coupling of the electrons with the electromagnetic field in free space by using the Peierls substitution, i.e., substituting in Eq.~(\ref{eq:H0}) via $t_{ij} \rightarrow t_{ij} e^{i \theta_{ij}}$, with the phase factor $\theta_{ij} = \frac{e}{\hbar} \int_{\bm{r}_{j}}^{\bm{r}_{i}} \bm{A} \bm{\cdot} d \bm{l}$. Here, $e = - |e|$ is the electron charge, $\hbar$ is the reduced Planck constant,  and $\bm{A}$ is the vector potential for describing the free space electromagnetic field. The coupling of the electrons with the electromagnetic field in the lowest order approximation can be obtained by expanding $\theta_{ij}$ in terms of $\bm{A}$ to the linear term,  given by
\begin{equation} \label{eq:Hint}
H_{\textrm{int}} = \sum_{\langle ij \rangle} \sum_{k} \sum_{\mu = x, y,z} M_{i j}^{k \mu} c_{i}^{\dag} c_{j} A_{\mu}(\bm{r}_{k}). 
\end{equation}
Here, the electron-photon coupling matrice is $M_{i j }^{k \mu} = i \frac{e}{2\hbar} t_{ij} (\bm{r}_{i} - \bm{r}_{j})_{\mu} (\delta_{ki} + \delta_{kj})$.

The EL from a SWCNT is considered by a typical two-terminal device under a bias voltage, as shown in Fig.~{\ref{fig:MD}}(a). The device consists of three parts. The left and right  leads are semi-infinite extensions of the pristine CNT. The central part is the conducting channel and it has a finite length. We take into account of the interaction of electrons with the EM field only in the central region. Upon applying a bias voltage, an electric current flows through the channel, and photons are excited and emitted due to the inelastic scattering of electrons interacting with the electromagnetic field. The cases that the channel contains a SV defect or a single SW defect are shown in Fig.~{\ref{fig:MD}}(b) and Fig.~{\ref{fig:MD}}(c) respectively. The SV defect is modeled by using a large onsite energy of $10^{6}$ eV for the vacant atom, and the SW defect is formed by rotating a C-C bond by 90 degrees. The effects of structure relaxations due to the defects are not considered in this paper. Also, we consider the NN hopping parameter $t_{ij} = t$ as a constant.

Radiation from the device is calculated using the nonequilibrium Green's function (NEGF) method based on our previous work~\cite{Wang2008,Wang2014,Zhang2020Angular,Zhang2020Farfield}. The important quantity is the local current-current correlation function. Its lesser component can be expressed in the random phase approximation as
\begin{equation} \label{eq:Pilocal}
\begin{split}
\Pi_{\mu \nu}^{<} & (\bm{r}_{i}, \bm{r}_{j}; \omega)  \\
 =& - i \hbar \int_{-\infty}^{\infty} \frac{dE}{2 \pi \hbar} \textrm{Tr} \big[ M^{i\mu} g^{<}(E) M^{j \nu} g^{>}(E- \hbar \omega) \big], 
\end{split}
\end{equation}
where, $\textrm{Tr}[\cdots]$ stands for trace over the electron degrees of freedom. The electron's lesser (greater) Green's function (GF) without coupling to the EM field is given by $g^{<(>)} = g^{r} \Sigma_{\textrm{leads}}^{<(>)} g^{a}$, with the retarded GF $g^{r}(E) = \big[ (E + i \eta)I -  H_{0} - \Sigma_{\textrm{leads}}^{r} \big]^{-1}$, and advanced GF $g^{a} = (g^{r})^{\dag}$. $I$ is the identity matrix, and $\eta$ is the GF infinitely small quantity. $\Sigma_{\textrm{leads}}^{r}$ is the total self-energy of the two  semi-infinite leads, which are calculated by using the recursive GF method~\cite{Nardelli1999}. Each lead is in equilibrium and follows the fluctuation-dissipation theorem, obeying the relation $\Sigma_{p}^{<} = - f_{p} (\Sigma_{p}^{r} - \Sigma_{p}^{a})$, with $p = \textrm{L}, \textrm{R}$ being the lead indices. $f_{p}(E, \mu_{p}) = 1/ \big[ \textrm{exp}(\frac{E - \mu_{p}}{k_{\textrm{B}} T_{p}}) + 1 \big]$ is the Fermi distribution function, $k_{\textrm{B}}$ is the Boltzmann constant, $\mu_{p}$ and $T_{p}$ are the temperature and chemical potential of the lead respectively. 

Using the monopole approximation and ignoring the screening effect on current fluctuations, the radiation power and rate of the photon counts (number of photons emitted per unit time) in the far field are given by~\cite{Zhang2020Angular}
\begin{eqnarray} 
P &=& - \int_{0}^{\infty} \frac{d \omega}{2 \pi} \frac{\hbar \omega^2}{3 \pi \varepsilon_{0} c^{3}} \sum_{\mu} \textrm{Im} \big[ \Pi_{\mu \mu}^{\textrm{tot},<}(\omega) \big],  \label{eq:Erad} \\
\frac{d N}{d t} &=& - \int_{0}^{\infty} \frac{d \omega}{2 \pi} \frac{\omega}{3 \pi \varepsilon_{0} c^{3}} \sum_{\mu} \textrm{Im} \big[ \Pi_{\mu \mu}^{\textrm{tot},<}(\omega) \big],  \label{eq:Nrad}\end{eqnarray}
where $\varepsilon_{0}$ is the vacuum permittivity, $c$ is the speed of light, and $\Pi_{\mu \mu}^{\textrm{tot},<} (\omega) = \sum_{ij} \Pi_{\mu \nu}^{<}(\bm{r}_{i}, \bm{r}_{j}; \omega)$ is the total current-current correlation function.

\begin{figure*} 
\centering
\includegraphics[width=16 cm]{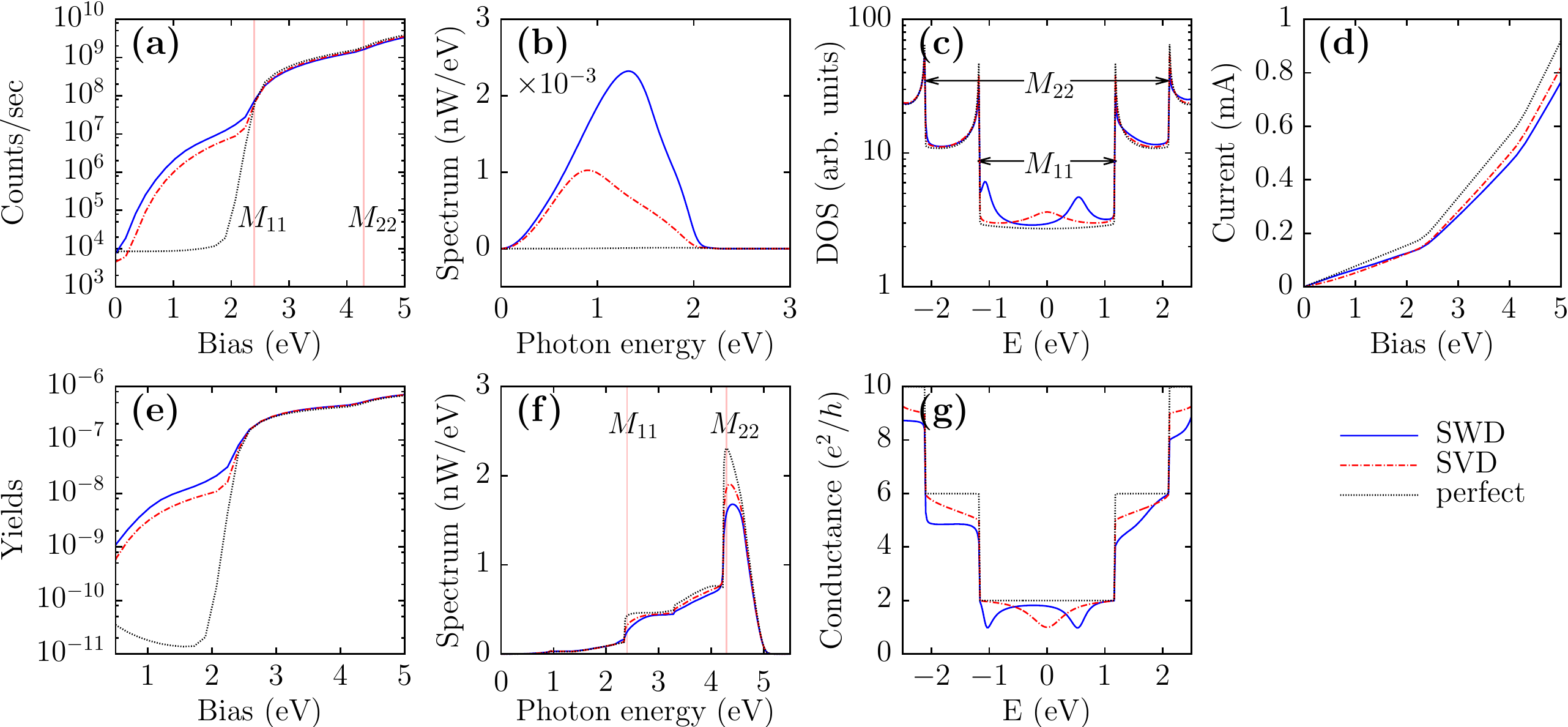}
\caption{Results for the two-terminal device of the CNT with chiral index (7, 7). Three curves are shown in each figure, for the CNTs with a single SW defect, a single SV defect, and the perfect one without defect respectively.  (a) Photon counts per second and (e) yields from the CNT under a bias voltage. (b) and (f) show the spectrum of the radiation power under the bias voltages 2.0~eV and 5.0~eV respectively. (c) The total density of states of the electrons in the channel. (g) The conductance and (d) the $I$-$V$ curve.
}
\label{fig:7v7}
\end{figure*}

%%%%%%%%%%%%%%%%%%%%%%%%%%%%%%%%%%%%%%%%%%%%%%
%\section{\label{sec:Apply} 
%%%%%%%%%%%%%%%%%%%%%%%%%%%%%%%%%%%%%%%%%%%%%%
{\emph{Results and discussions.}}
In the numerical calculation, we set the bias between the two leads to be symmetric, with $\mu_{\textrm{L}} = -\mu_{R} = V/2$. The C-C bond length is $a = 1.42 $~\AA. The NN hopping parameter is $t = 2.7$ eV~\cite{wilder1998electronic}. In this work, we consider the metallic CNTs to be armchair type and we do not consider spins of electrons. Temperatures for the two leads are both set to be $300$~K unless stated otherwise. Restricted by the computational cost, we use the central CNT channel with the length $l = 10\sqrt{3}a$. 

Firstly, we consider a typical metallic SWCNT with chiral index (7, 7) for the two-terminal device. We compare the results in each plot of Fig.~{\ref{fig:7v7}} for the cases that the CNT channel containing a single SW defect (SWD), a single SV defect (SVD), and no defect, respectively.  As shown in Fig.~{\ref{fig:7v7}}(a), for a  perfect CNT,  with the increasing of the bias voltage, the photon counts is very small and little changed in the low bias range until the onset of bias at about $2.4$~eV, which implies the opening of the $M_{11}$ transition between the two van Hove singularities shown by the DOS in Fig.~{\ref{fig:7v7}}(c). Below the onset of bias, thermal radiation is small but dominant for the perfect CNT. 

However, when the CNT channel contains a single SW defect or a single SV defect, the photon counts increase almost exponentially in the low bias range, the former is about two times as large as the latter. We plot in Fig.~{\ref{fig:7v7}}(b) and Fig.~{\ref{fig:7v7}}(f) the spectrum of the radiation power, which is defined as $S(\omega) = - \frac{\omega^2}{6 \pi^2 \varepsilon_{0} c^3} \sum_{\mu} \textrm{Im}\left[ \Pi_{\mu\mu}^{\textrm{tot},<} (\omega) \right]$ from the integrand of Eq.~(\ref{eq:Erad}),  setting the bias below and above the onset of bias for the $M_{11}$ transition respectively, with $V = 2.0$~eV and $V = 5.0$~eV. The spectrum in Fig.~{\ref{fig:7v7}}(b) shows that the average energy of the emitted photons from the CNT with a SW defect is larger than that from the CNT with a SV defect. The radiation spectrums for the perfect CNT and the defected CNTs under a large bias show little difference in Fig.~{\ref{fig:7v7}}(f), where the influence of the defects on the radiation is not obvious due to the strong transitions from high energy bands. To analyze the enhancement of EL in the low energy range, we plot in Fig.~{\ref{fig:7v7}}(c) the density of states of the CNT channel. There are extra peaks of the DOS in the low energy range induced by the defects. These localized states account for the EL in the low bias range. For the CNT with a SV defect, the localized state locates near the Fermi level, and for the case with a SW defect, the localized states are away from the Fermi level, close to the edge of the first pair of the van Hove singularities. Thus, the latter induces transitions to emit photons with higher energy on average than the former. Also, compared with that of the perfect CNT, the electric current in Fig.~\ref{fig:7v7}(d) decreases more significantly for the case with a SV defect than that with a SW defect in the low bias range. The localized states due to the defects in the low energy range reduce the conductance by one quantum unit, as shown in Fig.~{\ref{fig:7v7}}(g). The emission yield, i.e., the number of photons emitted per electron injected into the device channel, is an important quantity to characterize the emission efficiency of the device.  Here, the defects can enhance the counts and decrease the electric current, thus they enhance the yields of the EL, as shown in Fig.~{\ref{fig:7v7}}(e). The yields can reach the order of $10^{-7}$ in the high bias range, which is consistent experimental values~\cite{Freitag2004Hot}. 
 
 \begin{figure} 
\centering
\includegraphics[width=8 cm]{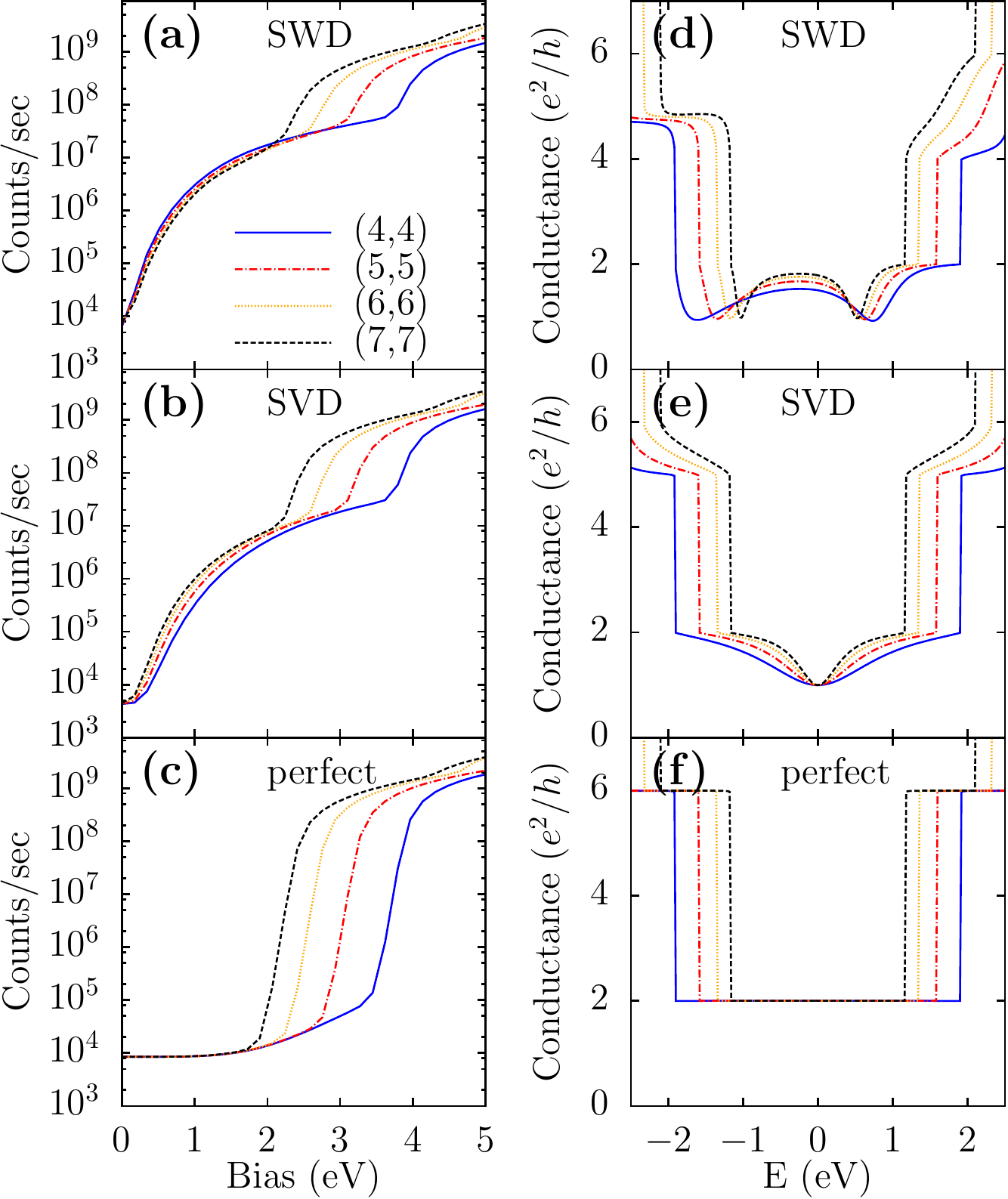}
\caption{Photon counts per second (left panel) and conductance (right panel) for the two-terminal device using CNTs with different diameter, with the chiral index (n,n) ranging from $n=4$ to n=7. } 
\label{fig:RadSize}
\end{figure}
 
In Fig.~{\ref{fig:RadSize}}, we discuss the influence of the diameter of the CNT to the EL. Specifically, We consider four different armchair CNTs with the chiral index (n,n) ranging from $n = 4$ to $n = 7$.  Their diameters are $d = 5.42$~\AA, $6.78$~\AA, $8.14$~\AA, and $9.49$~{\AA}, respectively.  There are distinct features for the ELs from the CNTs with different diameters, though the overall trends are  similar,  as shown in Fig.~{\ref{fig:RadSize}}(a)-(c). Firstly, plots for the counts from CNTs with different diameters formed some `bubbles' in the high bias range for the CNTs with defects in Fig.~{\ref{fig:RadSize}}(a) and Fig.~{\ref{fig:RadSize}}(b), and the perfect CNTs in Fig.~{\ref{fig:RadSize}}(c). This is due to that the transition energy corresponding to the $M_{11}$ gap decreases with the increasing of the tube diameter, which is shown from the conductance plots in Fig.~{\ref{fig:RadSize}}(d)-(f).  Secondly, the dependence of the counts on the tube diameter is very different for perfect CNTs and defected CNTs under low bias.  When the bias is smaller than the onsite bias of $M_{11}$ transition, photon counts are inversely proportional to the tube diameter for SW defected CNT, and they are proportional to the tube diameter for SV defected CNT, while they are nearly independent of the tube diameter for the perfect CNT, as shown in  Fig.~{\ref{fig:RadSize}}(a)-(c) respectively. In the low bias range, thermal radiation is dominant over the EL for the perfect CNTs. The energy dispersion relation in the low energy range accounts for the electron transport in the longitudinal direction along the tube axis and it is of little difference for the CNTs with different diameters, so they show little dependence of thermal radiation on the tube diameter  in Fig.~{\ref{fig:RadSize}}(c).  

\begin{figure} 
\centering
\includegraphics[width=8 cm]{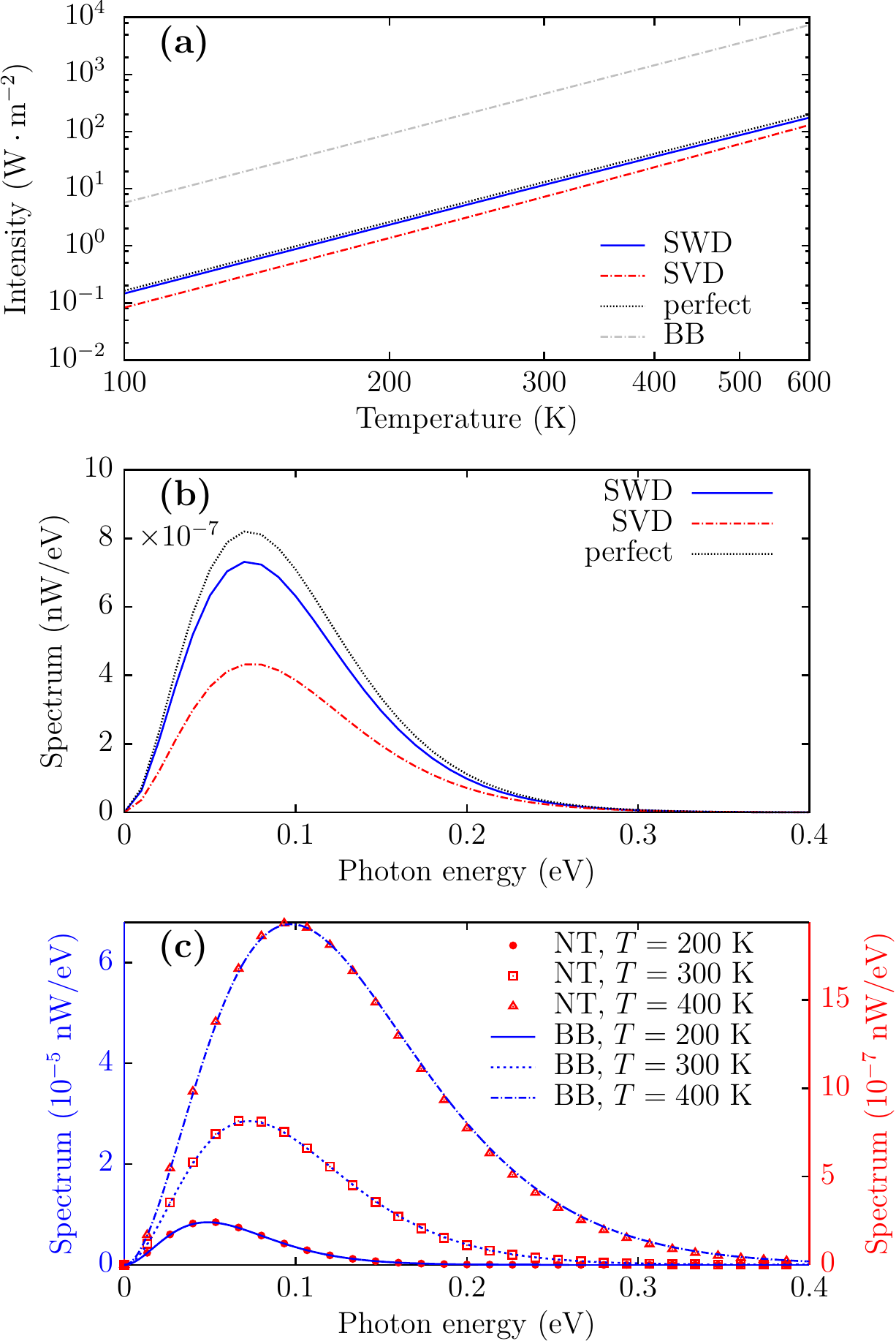}
\caption{Results of thermal radiation from the conducting channel of the two-terminal device using (7,7) CNT under zero bias. Temperatures of the two leads are the same. (a) Radiation intensity as a function of the temperature. (b) Spectrum of the radiation power with temperature $T = 300 \, \textrm{K}$. (c) The radiation spectrum using the perfect CNT under different temperatures are compared with the spectrum of the black-body radiation with the same area as the CNT channel. 
}
\label{fig:TherRad}
\end{figure}

We plot in Fig.~{\ref{fig:TherRad}} the thermal radiation from the CNT channel of the two-terminal device by turning off the bias voltage. Figure~{\ref{fig:TherRad}}(a) shows that the radiation power of the perfect CNT and the defected CNTs vs the temperature follows the $T^{4}$ scaling law as the black-body (BB) radiation. However, the intensity is about two orders of magnitude smaller than the black-body radiation. Figure~{\ref{fig:TherRad}}(b) shows the spectrum of the thermal radiation. Both the perfect and defected CNTs show black-body-like radiation, i.e., their spectrums fit well with that of the black-body radiation despite the magnitude is smaller. The fitting of the spectrum with the black-body radiation is shown in Fig.~{\ref{fig:TherRad}}(c), with the perfect CNT as an example. 

\begin{figure} 
\centering
\includegraphics[width=8 cm]{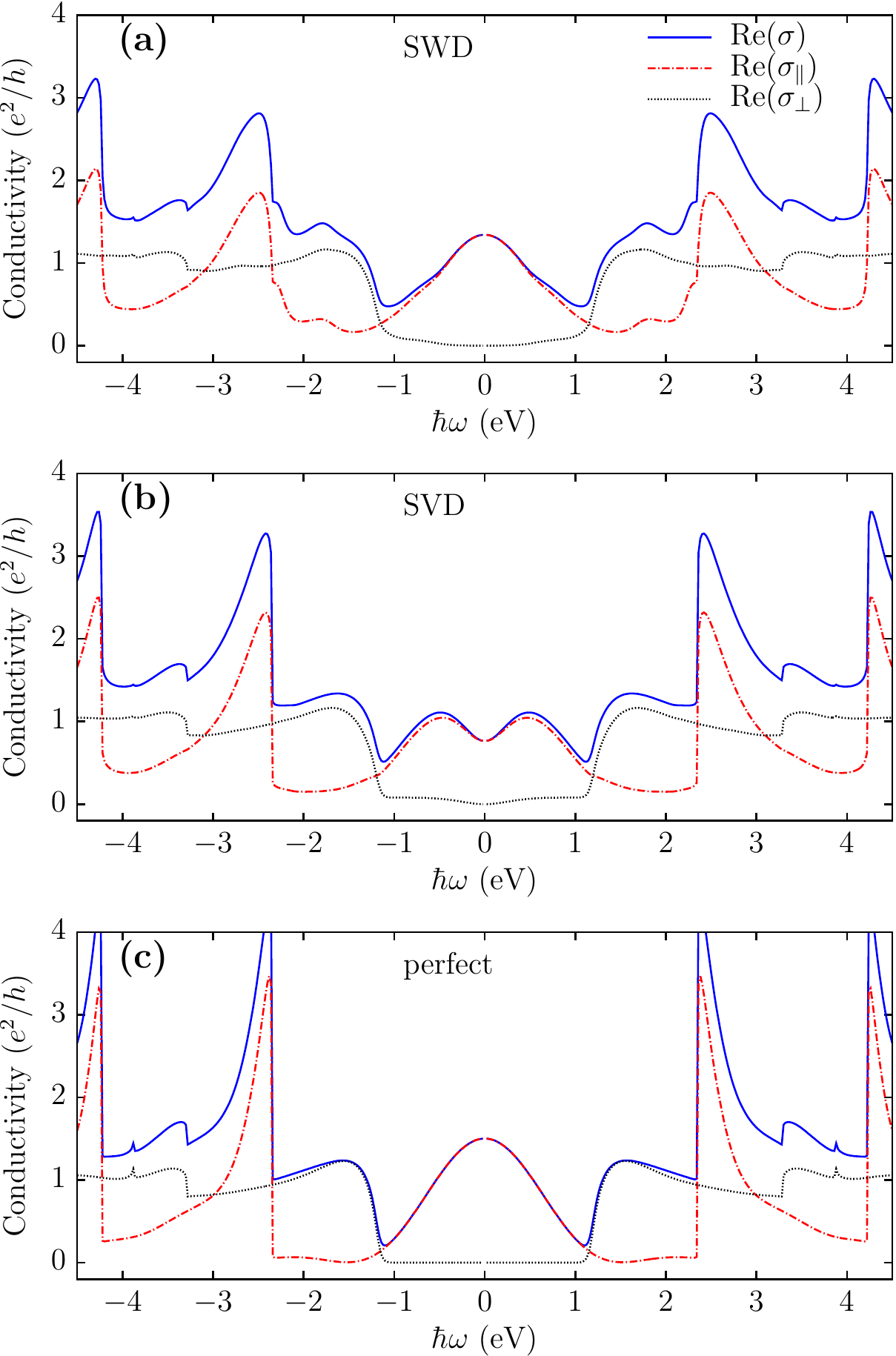}
\caption{Optical conductivity (real part) for the CNT channel of the device using the (7,7) CNTs at $T=300$ K. $\sigma$, $\sigma_{\parallel}$ and $\sigma_{\perp}$ are the total conductivity, the longitudinal part and perpendicular part respectively.  
}
\label{fig:Conductivity}
\end{figure}

Why the thermal radiations from the CNTs follows the black-body-like spectrum but they are much smaller than black-body radiation? To analysis this, we start from the general expression in Eq.~(\ref{eq:Erad}) to analysis the spectrum of the radiation power. The optical conductivity is related to the retarded component of the current-current correlation function as $\sigma(\omega) = \frac{i}{A \omega} \Pi^{r}(\omega)$, with $A= \pi d l$ the area of the central CNT channel. Using the fluctuation-dissipation relation in thermal equilibrium $\Pi^{<}(\omega) = i N_{B}(\omega) 2 \textrm{Im} [\Pi^{r}(\omega)]$, with $N_{B}(\omega)$ the Bose distribution function, we can write from Eq.~(\ref{eq:Erad}) the spectrum of the radiation power in thermal equilibrium as 
\begin{equation}
S(\omega) = \frac{A \omega^3}{3 \pi^2 \varepsilon_{0} c^3} \textrm{Re} [\sigma^{\textrm{tot}}(\omega)] N_{B} (\omega).
\label{eq:Spectrum}
\end{equation}
Here we use the notation $\sigma^{\textrm{tot}}(\omega) = \sum_{\mu = x, y, z}\sigma_{\mu \mu}(\omega)$. The optical conductivity is calculated by
\begin{equation}
\begin{split}
\sigma_{\mu \nu}&(\omega)   \\
=& \frac{1}{A} \int_{-\infty}^{\infty} \frac{dE}{2 \pi \omega} \sum_{ij} \textrm{Tr} \big[ M^{i\mu} g^{r}(E) M^{j \nu} g^{<}(E- \hbar \omega)  \\
+& M^{i\mu} g^{<}(E) M^{j \nu} g^{a}(E- \hbar \omega) \big]. 
\end{split}
\end{equation}
The longitudinal and transverse components of the conductivity are given by $\sigma_{\parallel} = \sigma_{zz}$, and $\sigma_{\perp} = \sigma_{xx} + \sigma_{yy}$ respectively. The spectrum of black-body radiation with the same area is $S_{\textrm{BB}} (\omega) = \frac{A \omega^3}{4 \pi^2 c^2} N_{B}(\omega)$. Comparing it with Eq.~(\ref{eq:Spectrum}), we conclude that the strict condition for the shape of the radiation spectrum of a metallic material to fit with that of the black-body radiation is that the real part of the conductivity should be a constant in the energy range of thermal excitation. The black-body-like spectrum in Fig.~{\ref{fig:TherRad}}(b) is determined to a large extent by the intrinsic nature of thermal equilibrium fluctuation, i.e., the factor $\omega^3 N_{B}(\omega)$ in Eq.~(\ref{eq:Spectrum}), despite that the real parts of the conductivities for the channels using the perfect and defected CNTs in Fig.~{\ref{fig:Conductivity}} are not strictly constant in the energy range of thermal excitation. For both the perfect and defected CNTs, the total conductivity is mainly contributed by the longitudinal component in the low energy range, noting that the photon energy from thermal radiation is smaller than $0.5$~eV, while the transverse component is only significant in the high energy range.  Thus, thermal radiation is much smaller than the black-body radiation due to the constraint of the excitations in the circular direction of the CNTs. Finally, a quantitative analysis of the change of the magnitude of the spectrum due to the defects of CNTs in Fig.~{\ref{fig:TherRad}}(b) can be attributed to the decrease of the conductivity by the defects shown in Fig.~{\ref{fig:Conductivity}}. 

%%%%%%%%%%%%%%%%%%%%%%%%%%%%%%%%%%%%%%%%%%%%%%
%\section{\label{sec:SUM} summary}
%%%%%%%%%%%%%%%%%%%%%%%%%%%%%%%%%%%%%%%%%%%%%%

%%\textcolor{red}{Comment: reference people name P. Avouris should be corrected to be Ph. Avouris?}

{\emph{Conclusion.}}
Using the NEGF method, we study the EL and thermal radiation from metallic SWCNTs with defects in the ballistic transport regime based on a tight-binding model. We find that both the SV defect and SW defect can enhance the EL, which increases exponentially in the low bias range,  while for the perfect nanotube only thermal radiation contributes and the EL can be neglected. The enhancement of radiation due to the defects is not obvious in the high bias range, where strong radiation due to transitions between high energy bands becomes dominant. The enhancement of the EL and the diameter of the CNT have a positive correlation in the presence of a SW defect, while for the CNT with a SV defect they have a negative correlation. Due to confinement of thermal excitation in the transverse direction, the intensity of the thermal radiation is much smaller than that of the black-body radiation and it is independent of the nanotube diameter. Defects can reduce the optical conductivity of the CNT, and they reduce the thermal radiation. This reducing effect is more significant for the CNT with a SV defect than that with a SW defect.

\begin{acknowledgments}
We acknowledge the support by MOE tier 2 Grant No. R-144-000-411-112 and FRC Grant No. R-144-000-402-114. 
\end{acknowledgments}

\bibliographystyle{apsrev4-1}
\bibliography{RadCNT}

\end{document}